\documentclass[twocolumn,showpacs,superscriptaddress,amsmath,amssymb,aps,pra]{revtex4-1}
\usepackage{bm}
\usepackage{graphicx}
\usepackage{dcolumn}
\usepackage{amsmath,txfonts}
\usepackage{epstopdf}
\usepackage[mathlines]{lineno}
\usepackage{color}
\usepackage[colorlinks=true,linkcolor=blue,citecolor=blue]{hyperref}
\usepackage[normalem]{ulem}
\usepackage{makecell}
\usepackage{tabularx}

\begin{document}

\title{Nonreciprocal generation of Bell states via cooperative dissipation in a qubit-magnon system}

\author{Fei Gao}
\affiliation{School of Physics, Hangzhou Normal University, Hangzhou 311121, China}

\author{Xiang-Jing Geng}
\affiliation{School of Physics, Hangzhou Normal University, Hangzhou 311121, China}

\author{Wei Feng}
\affiliation{School of Physics, Hangzhou Normal University, Hangzhou 311121, China}

\author{Lijiong Shen}
\affiliation{School of Physics, Hangzhou Normal University, Hangzhou 311121, China}

\author{Guo-Qiang Zhang}
\email{zhangguoqiang@hznu.edu.cn}
\affiliation{School of Physics, Hangzhou Normal University, Hangzhou 311121, China}
\affiliation{Zhejiang Key Laboratory of Micro-Nano Quantum Chips and Quantum Control, Hangzhou 310027, China}

\begin{abstract}
Directionally generating and controlling quantum states is critical in chiral quantum technologies. Here, we propose a scheme for the nonreciprocal generation of Bell states via cooperative dissipation in a qubit-magnon system, consisting of a superconducting transmon qubit and a yttrium iron garnet sphere subject to a bias magnetic field. The cooperative dissipation between them is induced by coupling both to a common transmission-line waveguide. Due to this cooperative dissipation, the system possesses a dark eigenstate within the single-excitation subspace. Under appropriate parameter conditions, the dark eigenstate reduces to a specific Bell state. With the assistance of a drive pulse, the system can be prepared in this Bell state (i.e., the dark eigenstate). In particular, we find that the dark eigenstate corresponds to different Bell states under positive ($+\mathbf{B}_0$) and negative ($-\mathbf{B}_0$) bias fields. Consequently, nonreciprocal Bell-state generation is achieved: $+\mathbf{B}_0$ yields one Bell state, while $-\mathbf{B}_0$ yields the other. Moreover, we define a fidelity-based nonreciprocal rate to quantify this nonreciprocity. Our work establishes cooperative dissipation as a versatile resource for nonreciprocal entangled-state generation, opening a promising route toward chiral quantum technologies.
\end{abstract}

\date{\today}

\maketitle

\section{Introduction}

Owing to their promising applications in quantum technologies, chiral quantum effects and nonreciprocal quantum effects have attracted considerable attention~\cite{Lodahl17,SuarezForero25,Barzanjeh26,Jiao26}. Typical examples include directional spontaneous emission~\cite{Mitsch14,Sollner15,Coles16,Germanis25}, chiral-enabled quantum gate~\cite{Ralph15,Schrinski22,Levy-Yeyati25}, one-way quantum amplification~\cite{Metelmann15,Wanjura20,Pucher22,Hu23}, nonreciprocal quantum synchronization~\cite{Lai25,Nadolny25,Chabar26}, nonreciprocal superradiant phase transition~\cite{Chiacchio23,Zhu24,Xu-Zhang24,Zhang26}, nonreciprocal photon and phonon blockade~\cite{Huang18,Shen20,Zheng25,Yao22}, and nonreciprocal quantum entanglement~\cite{Jiao20,Chen23,Chakraborty23,Bin24,Chen24,Taravati25}. Notably, it is highly appealing to combine chirality and nonreciprocity with quantum state transfer and preparation, as this enables directional control of quantum information.

Early studies focused mainly on perfectly transferring a single-qubit quantum state from one end of a coupled spin chain to the other~\cite{Bose03,Christandl04,Li-Ma18}. However, in such schemes, the sender and receiver are fixed, which hinders system scalability. Subsequently, the study of single-qubit state transfer was extended to multi-node quantum networks~\cite{Chudzicki10,Pemberton-Ross11,Roushan17,Wang-Song19,Feng25,Lu25}. Recently, both chiral and nonreciprocal schemes for generating Schr\"{o}dinger cat states have been proposed, based on the Sagnac-Fizeau effect in a spinning resonator~\cite{Liu25} and the topological phase transition in a one-dimensional microcavity array~\cite{Li24}, respectively. Extending such chiral control from single-mode states to entangled states, chiral Bell-state transfer and preparation has been investigated via dynamically encircling a second-order exceptional point~\cite{Khandelwal24,Wu26,Xiao26}. Nevertheless, the nonreciprocal preparation of Bell states remains an open problem.

Among the diverse physical platforms for exploring chiral and nonreciprocal quantum effects, magnon-based hybrid systems hold a prominent position owing to their excellent tunability and design flexibility~\cite{Lachance-Quirion19,Yuan22,Rameshti22}. Experimentally, researchers introduced cooperative dissipation of the cavity and magnon modes by coupling a microwave cavity and a yttrium iron garnet (YIG) sphere to a common waveguide, successfully observing nonreciprocal microwave transmission by reversing either the microwave propagation direction~\cite{Wang19,Zhao20,Kim24} or the bias magnetic field~\cite{Qian20,Yao25}. In these systems, the nonreciprocity arises predominantly from the intrinsic asymmetric magnetic permeability tensor of the YIG sphere~\cite{So95,Zhang21}. Building upon this mechanism, subsequent experiments have demonstrated nonreciprocal control of the speed of light~\cite{Yao25}, chiral-selected circular transmission~\cite{Shi21}, and unidirectional microwave amplification~\cite{Wang-Qian23}, while theoretical studies have proposed nonreciprocal magnon blockade~\cite{Wang22}, one-way steering~\cite{Guan22}, nonreciprocal phonon laser~\cite{Li25}, and so on. This naturally raises an intriguing question: can such a cooperative-dissipation-based nonreciprocal mechanism be further extended to achieve the nonreciprocal preparation of quantum entangled states?

In this paper, we develop a scheme to achieve nonreciprocal generation of Bell states via cooperative dissipation in a hybrid quantum system. The considered system consists of a superconducting (SC) transmon qubit and a ferrimagnetic YIG sphere magnetized to saturation by a bias magnetic field, both coupled to a transmission-line waveguide. Here the waveguide serves as a common reservoir, which induces the cooperative dissipation between the qubit and the magnon mode in the YIG sphere. By analyzing the effective non-Hermitian Hamiltonian of the qubit-magnon system, we identify a dark eigenstate within the single-excitation subspace. With appropriate system parameters, the dark eigenstate can be mapped onto a specific Bell state. Under a suitable drive pulse, the system evolves from its ground state into this dark state (i.e., the target Bell state), thereby achieving Bell-state preparation. Interestingly, we find that applying positive ($+\mathbf{B}_0$) and negative ($-\mathbf{B}_0$) bias fields yields different Bell states (i.e., one Bell state for $+\mathbf{B}_0$ and the other for $-\mathbf{B}_0$), thus realizing nonreciprocal Bell-state generation. This nonreciprocity originates from the asymmetric magnetic permeability tensor of the YIG sphere~\cite{So95,Zhang21}, which leads to different cooperative jump operators upon magnetic field reversal [cf.~Eq.~(\ref{jump-operator})]. Furthermore, a fidelity-based nonreciprocal rate is introduced to quantitatively characterize the nonreciprocity. To the best of our knowledge, this is the first theoretical proposal for nonreciprocal Bell-state generation, which offers promising applications in chiral quantum technologies~\cite{Liu25,Li24,Khandelwal24,Wu26,Xiao26}.

The remainder of this paper is organized as follows. In Sec.~\ref{model}, we describe the qubit-magnon system and identify the dark eigenstate within the single-excitation subspace. In Sec.~\ref{GBS}, we demonstrate the nonreciprocal generation of Bell states, analyze the impact of system decoherence on our scheme, and introduce a fidelity-based nonreciprocal rate to quantify the degree of nonreciprocity. Finally, we discuss the experimental feasibility of the proposed scheme and give a brief summary in Sec.~\ref{Discussion}.

\begin{figure}
\includegraphics[width=0.49\textwidth]{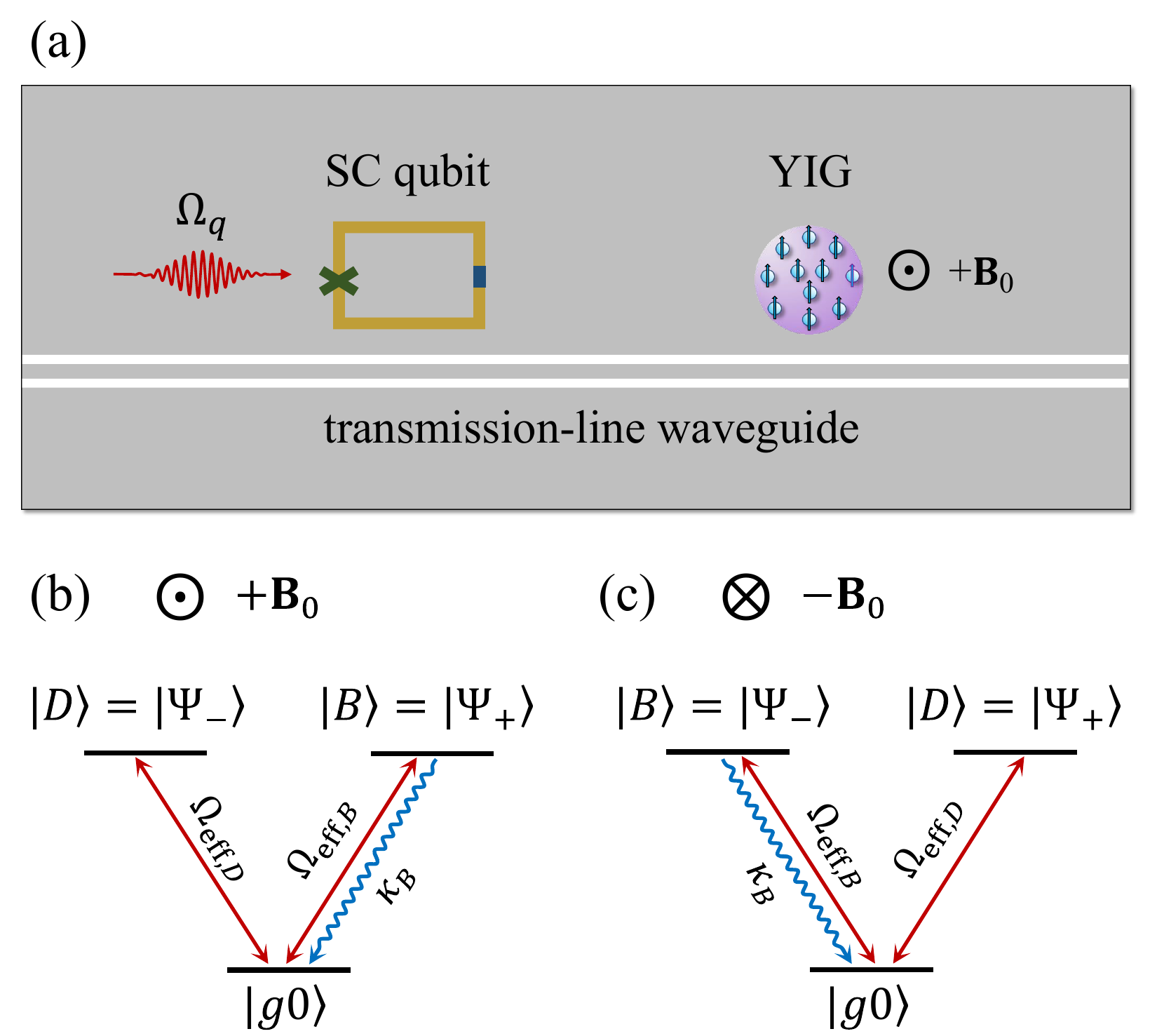}
\caption{(a) Schematic diagram of the proposed hybrid quantum system for nonreciprocally generating Bell states. The transmission-line waveguide induces cooperative dissipation between the SC transmon qubit and the magnon mode in the ferrimagnetic YIG sphere, which is magnetized to saturation by an externally applied magnetic field $+\mathbf{B}_0$. To generate the Bell state $|\Psi_-\rangle$, a coherent microwave field is applied to drive the SC qubit. (b),(c) Energy-level diagrams of the qubit-magnon system under positive ($+\mathbf{B}_0$) and negative ($-\mathbf{B}_0$) bias magnetic fields, respectively. The drive field pumps the system from the ground state $|g0\rangle$ to the bright state $|B\rangle$ and the dark state $|D\rangle$. For a positive (negative) magnetic field, $|D\rangle=|\Psi_-\rangle$ and $|B\rangle=|\Psi_+\rangle$ ($|B\rangle=|\Psi_-\rangle$ and $|D\rangle=|\Psi_+\rangle$). The bright state $|B\rangle$ superradiantly decays to the ground state $|g0\rangle$, whereas the dark state $|D\rangle$ remains decoupled from the waveguide.}
\label{fig1}
\end{figure}

\section{Model}\label{model}

As shown in Fig.~\ref{fig1}(a), the hybrid quantum system under consideration consists of an SC transmon qubit and a ferrimagnetic YIG sphere, which are both coupled to a common transmission-line waveguide. The Hamiltonian of the SC qubit is $H_q=\omega_q \sigma_+ \sigma_-$, where $\omega_q$ is the qubit transition frequency, and $\sigma_{\pm}=(\sigma_x \pm i\sigma_y)/2$ are the raising and lowering operators expressed in terms of the Pauli operators $\sigma_x$, $\sigma_y$, and $\sigma_z$. When the YIG sphere is uniformly magnetized to saturation, only the Kittel mode can be excited, and its Hamiltonian is $H_m=\omega_m m^\dag m$, with the creation (annihilation) operator $m^\dag$ ($m$) of the magnon mode at frequency $\omega_m$. To achieve nonreciprocal Bell-state generation, we apply a coherent microwave pulse, with frequency $\omega_d$ and duration $t_0$, to drive the qubit. The corresponding drive Hamiltonian takes the form $H_d=\theta(t_0-t)\Omega_d (\sigma_+ e^{-i\omega_d t}+\sigma_- e^{i\omega_d t})$, where $\theta(t_0-t)$ is the Heaviside function, and $\Omega_d$ is the drive amplitude. The total Hamiltonian of the hybrid system, $\mathcal{H}=H_q +H_m +H_d$, can then be written as
\begin{eqnarray}\label{Hamiltonian}
\mathcal{H}=\omega_q \sigma_+ \sigma_- + \omega_m m^\dag m
             +\theta(t_0-t)\Omega_d (\sigma_+ e^{-i\omega_d t}+\sigma_- e^{i\omega_d t}).~~~~
\end{eqnarray}
In our scheme, the waveguide acts as a shared reservoir. Under the Born-Markovian approximation, the influence of the waveguide on the qubit-magnon system is captured by a standard Lindblad dissipator~\cite{Lalumiere13}. Including the waveguide contribution, the evolution of the density operator $\rho$ of the qubit-magnon system is governed by~\cite{Wang19,Wang22,Guan22,Li25}
\begin{equation}\label{master1}
\dot{\rho}=-i[\mathcal{H},\rho]+\Gamma\mathcal{L}[\mathcal{O}]\rho,
\end{equation}
where the dissipator is defined as
\begin{equation}\label{superoperator}
\mathcal{L}[\mathcal{O}]\rho=2\mathcal{O}\rho \mathcal{O}^\dag-\mathcal{O}^\dag \mathcal{O}\rho-\rho\mathcal{O}^\dag \mathcal{O}.
\end{equation}
This dissipator describes the cooperative dissipation of the qubit and the magnon mode into the waveguide, characterized by the cooperative decay rate $\Gamma$. The cooperative jump operator $\mathcal{O}$ is a linear combination of the qubit lowering operator $\sigma_-$ and the magnon annihilation operator $m$. Owing to the asymmetric magnetic permeability tensor of the YIG sphere, the jump operator $\mathcal{O}$ becomes nonreciprocal under both the positive ($+\mathbf{B}_0$) and negative ($-\mathbf{B}_0$) bias magnetic fields. Here, $-\mathbf{B}_0$ is in the opposite direction to $+\mathbf{B}_0$. For convenience, $\mathcal{O}$ can be written as~\cite{Wang19,Wang22,Guan22,Li25}
\begin{equation}\label{jump-operator}
\mathcal{O}=\mu\sigma_- +e^{i\Theta}\nu m,
\end{equation}
where the real coefficients $\mu$ and $\nu$ denote the coupling strengths of the qubit and magnon mode to the waveguide, respectively, and the phase $\Theta$ is determined by the direction of the bias field~\cite{Qian20,Yao25}. For a positive bias field, $\Theta=0$ and the jump operator reduces to $\mathcal{O}=\mu\sigma_- +\nu m$. For a negative bias field,
$\Theta=\pi$ and the jump operator becomes $\mathcal{O}=\mu\sigma_- -\nu m$. This nonreciprocity of the jump operator consequently leads to the nonreciprocal generation of Bell states (cf. Sec.~\ref{GBS}).

\renewcommand\tabcolsep{8.9pt}
\begin{table}
	\centering
	\scriptsize
	\caption{Dark state $|D\rangle$ and bright state $|B\rangle$ from Eq.~(\ref{dark-state}) for positive ($+\mathbf{B}_0$) and negative ($-\mathbf{B}_0$) bias magnetic fields, with $\mu=\nu=1$.}
	\label{tab:notations}
    \small
	\begin{tabular}{cccc}
        \hline
		\makecell[cc]{Bias magnetic field}  & \makecell[cc]{Phase} & \makecell[cc]{Dark state}& \makecell[cc]{Bright state}\\
		\hline
		\makecell[cc]{Positive ($+\mathbf{B}_0$)}  &\makecell[cc]{$\Theta=0$}     & \makecell[cc]{$|D\rangle=|\Psi_-\rangle$}
                                                                     & \makecell[cc]{$|B\rangle=|\Psi_+\rangle$}\\
        \hline
		\makecell[cc]{Negative ($-\mathbf{B}_0$)}  & \makecell[cc]{$\Theta=\pi$}& \makecell[cc]{$|D\rangle=|\Psi_+\rangle$} &
                                 \makecell[cc]{$|B\rangle=|\Psi_-\rangle$} \\
		\hline
	\end{tabular}
\label{Table1}
\end{table}

To analyze the eigenstates of the system in the absence of drive field (i.e., $\Omega_d=0$), we rewrite Eq.~(\ref{master1}) as
\begin{equation}\label{master2}
	\dot{\rho}=-i\left(\mathcal{H}_{\rm eff}\rho-\rho\mathcal{H}_{\rm eff}^{\dagger}\right)
                      +2\Gamma\mathcal{O}\rho\mathcal{O}^{\dagger},
\end{equation}
where $\mathcal{H}_{\rm eff}$ is the effective non-Hermitian Hamiltonian of the qubit-magnon system,
\begin{equation}\label{Heff}
\mathcal{H}_{\rm eff}=\omega_q \sigma_+ \sigma_- + \omega_m m^\dag m-i\Gamma\mathcal{O}^{\dagger}\mathcal{O}.
\end{equation}
When the qubit and magnon are resonant (i.e., $\omega_{q}=\omega_{m}=\omega_{0}$), $\mathcal{H}_{\rm eff}$ has two orthogonal eigenvectors in the one-excitation subspace spanned by $\{|e0\rangle, |g1\rangle\}$. Here $|g\rangle$ ($|e\rangle$) denotes the ground (excited) state of the qubit, and $|n\rangle$ ($n=0,\,1$) is the Fock state of the magnon mode with $n$ excitations. The two eigenvectors are
\begin{equation}\label{dark-state}
\begin{split}
|D\rangle=\frac{1}{\sqrt{\mu^{2}+\nu^{2}}}(\mu|e0\rangle-\nu e^{-i\Theta}|g1\rangle),\\
|B\rangle=\frac{1}{\sqrt{\mu^{2}+\nu^{2}}}(\nu|e0\rangle+\mu e^{-i\Theta}|g1\rangle),
\end{split}
\end{equation}
with eigenvalues $E_{D}=\omega_{0}$ and $E_{B}=\omega_{0}-i(\mu^{2}+\nu^{2})\Gamma$, respectively. The dark state $|D\rangle$ has zero decay rate and is therefore stable, i.e., it is decoupled from the waveguide. In contrast, the bright state $|B\rangle$ is superradiant and decays at a rate $\kappa_B=(\mu^2 + \nu^2)\Gamma$. For $\mu=\nu=1$, the two eigenvectors simplify to
\begin{equation}\label{}
\begin{split}
|D\rangle=\frac{1}{\sqrt{2}}(|e0\rangle-e^{-i\Theta}|g1\rangle),\\
|B\rangle=\frac{1}{\sqrt{2}}(|e0\rangle+e^{-i\Theta}|g1\rangle).
\end{split}
\end{equation}
Unless otherwise stated, we henceforth adopt this case as the default for the remainder of this paper.
When $\Theta=0$, one obtains $|D\rangle=|\Psi_{-}\rangle$ and $|B\rangle = |\Psi_{+}\rangle$; for $\Theta=\pi$, the two states are interchanged: $|D\rangle=|\Psi_{+}\rangle$ and $|B\rangle=|\Psi_{-}\rangle$. Here $|\Psi_{\pm}\rangle$ are the Bell states,
\begin{equation}\label{bell-state}
|\Psi_{\pm}\rangle=\frac{1}{\sqrt{2}}(|e0\rangle\pm|g1\rangle).
\end{equation}
For the reader's convenience, the dark and bright states for $\Theta=0$ and $\Theta=\pi$ are summarized in Table~\ref{Table1}.

\begin{figure}
\includegraphics[width=0.48\textwidth]{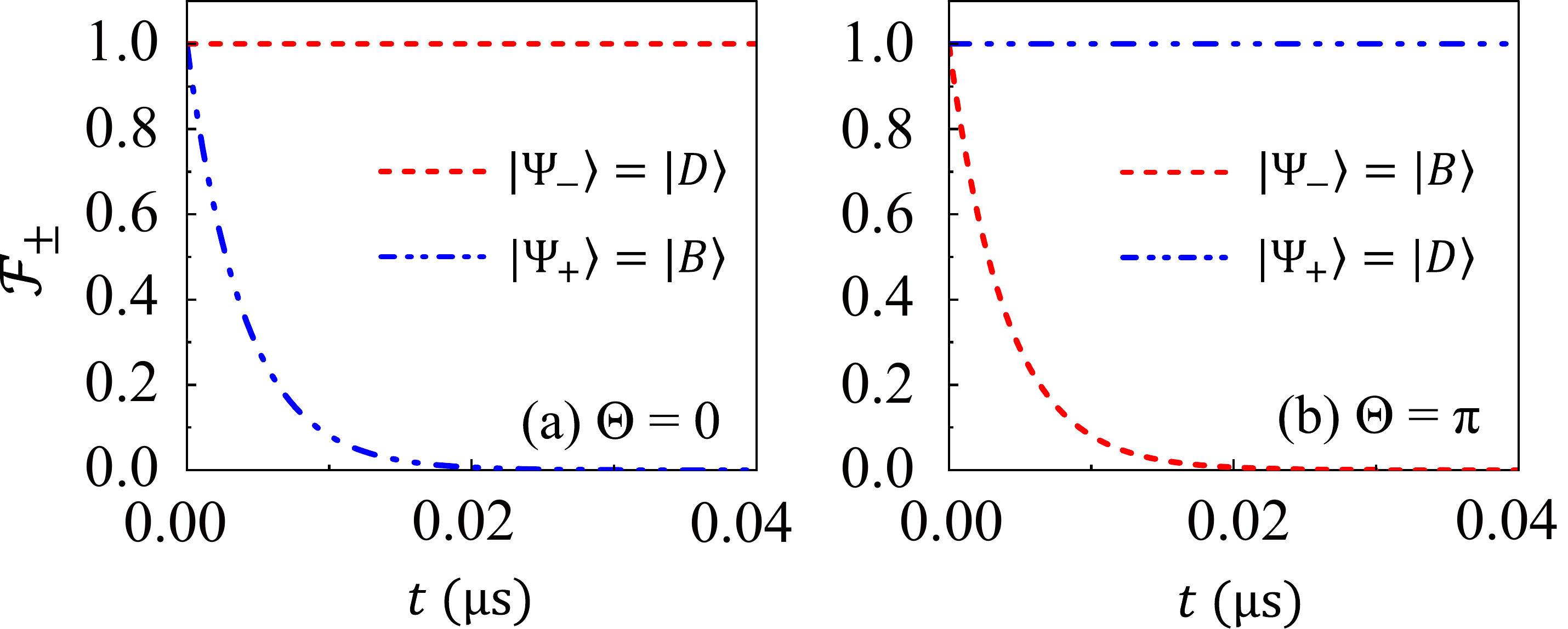}
\caption{Time evolution of the fidelity $\mathcal{F}_{X}={\rm Tr}(\rho |X\rangle\langle X|)$  for initial states $|X\rangle=|D\rangle$ and $|X\rangle=|B\rangle$. (a) $\Theta=0$: $|\Psi_{-}\rangle=|D\rangle$ (red dashed) and $|\Psi_{+}\rangle=|B\rangle$ (blue dash-dot-dot). (b) $\Theta=\pi$: $|\Psi_{-}\rangle=|B\rangle$ (red dashed) and  $|\Psi_{+}\rangle=|D\rangle$ (blue dash-dot-dot). Simulations are based on the master equation in Eq.~(\ref{master1}) with parameters $\omega_m/2\pi=\omega_q/2\pi=5$~GHz, $\Omega_d=0$, $\Gamma/2\pi=10$~MHz, and $\mu=\nu=1$.}
\label{fig2}
\end{figure}

To validate our theoretical analysis of the stability of the eigenstates $|B\rangle$ and $|D\rangle$, we perform numerical simulations of the system evolution by solving the master equation in Eq.~(\ref{master1}). As shown in Fig.~\ref{fig2}, when the initial state is chosen as $\rho_0=|D\rangle\langle D|$, the fidelity $\mathcal{F}_D={\rm Tr}(\rho |D\rangle \langle D|)$ remains identically 1 at all times, confirming that $|D\rangle$ is stable [see the red dashed curve in Fig.~\ref{fig2}(a) and the blue dash-dot-dot curve in Fig.~\ref{fig2}(b)]. When the initial state is $\rho_0=|B\rangle\langle B|$, by contrast, the fidelity $\mathcal{F}_B={\rm Tr}(\rho |B\rangle \langle B|)$ decays rapidly from 1 to 0, indicating that the system relaxes from the bright state $|B\rangle$ to the ground state $|g0\rangle$ [see the blue dash-dot-dot curve in Fig.~\ref{fig2}(a) and the red dashed curve in Fig.~\ref{fig2}(b)].

\section{Nonreciprocal generation of Bell states}\label{GBS}

\begin{figure}
\includegraphics[width=0.48\textwidth]{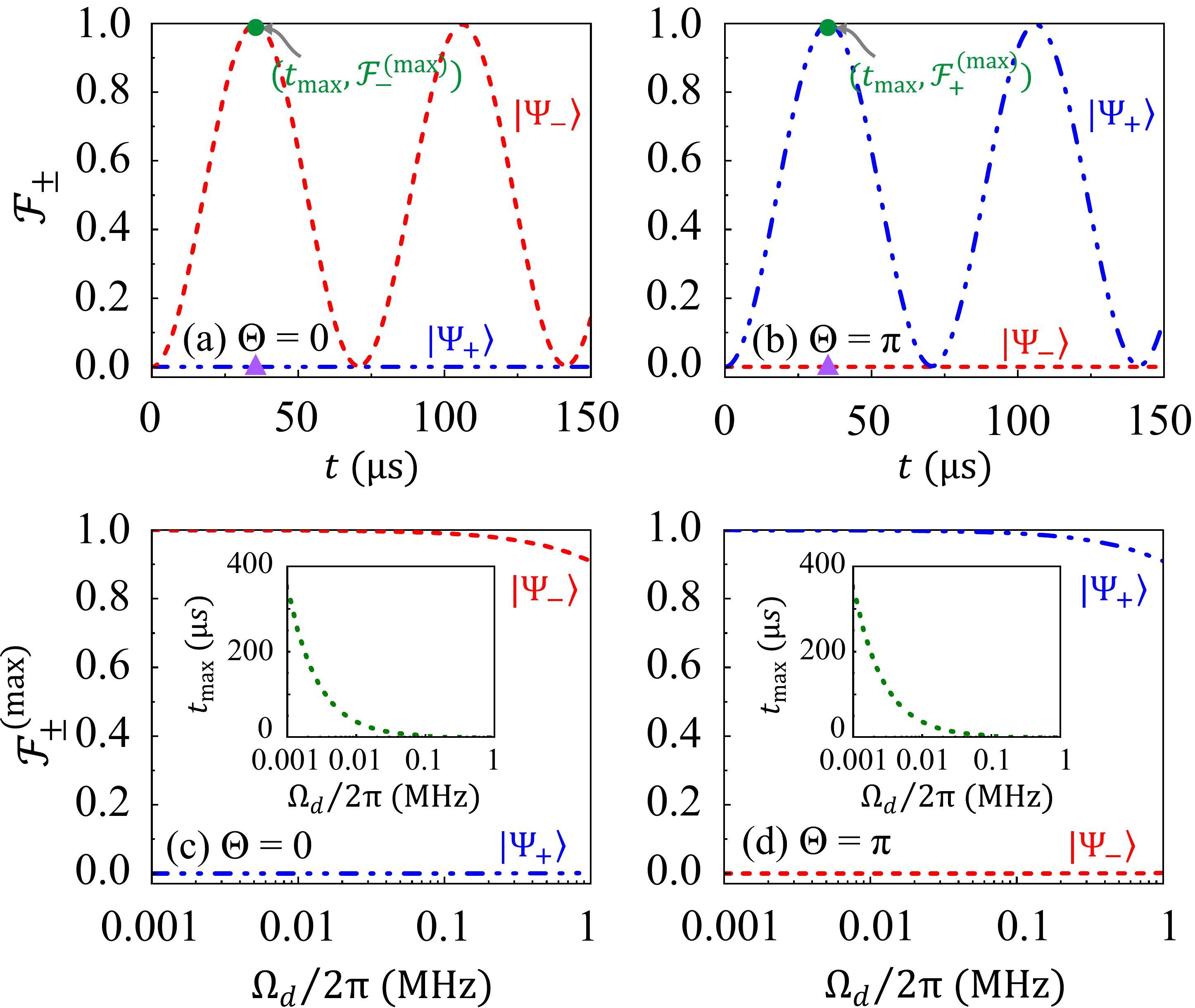}
\caption{(a),(b) Time evolution of the fidelity $\mathcal{F}_{\pm}$ of Bell states $|\Psi_{\pm}\rangle$ for (a) $\Theta=0$ and (b) $\Theta=\pi$, respectively, with $\Omega_d/2\pi=0.01$~MHz and $t_{0}=+\infty$. The green dots indicate the maximum fidelity $\mathcal{F}_{\pm}^{({\rm max})} = 0.990$ at $t_{\rm max} = 35.34$~$\mu$s. (c),(d) Maximum fidelity $\mathcal{F}_{\pm}^{(\rm max)}$ versus the Rabi frequency $\Omega_d/2\pi$ for (c) $\Theta=0$ and (d) $\Theta=\pi$, respectively,  where the corresponding time $t_{\rm max}$ is shown in the insets. In (c), $\mathcal{F}_+^{(\rm max)}$ corresponds to the value of $\mathcal{F}_+$ at $t=t_{\rm max}$ [see the purple triangle in (a)], while in (d), $\mathcal{F}_{-}^{({\rm max})}$ corresponds to the value of $\mathcal{F}_-$ at $t=t_{\rm max}$ [see the purple triangle in (b)]. In all panels, red dashed curves represent $|\Psi_{-}\rangle$, and blue dash-dot-dot curves represent $|\Psi_{+}\rangle$. These results in (a)--(d) are obtained using the master equation (\ref{master1}) with the Hamiltonian in Eq.~(\ref{Hamiltonian}) for the initial state $|g0\rangle$ and parameters $\Gamma/2\pi=10\ \mathrm{MHz}$, and $\omega_m/2\pi=\omega_q/2\pi=\omega_d/2\pi=5\ \mathrm{GHz}$.}
\label{fig3}
\end{figure}

The dependence of the jump operator on the direction of the bias magnetic field [see Eq.~(\ref{jump-operator})] makes it possible to prepare Bell states in a nonreciprocal manner. Starting from the same initial state, one Bell state can be prepared for $+\mathbf{B}_0$ (i.e., $\Theta=0$), whereas the other Bell state is obtained for $-\mathbf{B}_0$ (i.e., $\Theta=\pi$). Initially, the system is prepared in the ground state $|g0\rangle$. Upon driving the qubit, the qubit is excited and the system evolves to $|e0\rangle$. Due to $|e0\rangle = \big(|\Psi_-\rangle + |\Psi_+\rangle\big)/\sqrt{2}$, this process corresponds to the system evolving from the ground state $|g0\rangle$ into the Bell states $|\Psi_{-}\rangle$ and $|\Psi_+\rangle$ [cf. Figs.~\ref{fig1}(b) and \ref{fig1}(c)]. When $\Theta=0$, $|\Psi_{-}\rangle$ ($=|D\rangle$) is the dark state and $|\Psi_{+}\rangle$ ($=|B\rangle$) is the bright state. Thus, the component of $|\Psi_{-}\rangle$ is stable, while the component of $|\Psi_{+}\rangle$ decays to the ground state $|g0\rangle$ at a rate $\kappa_B=2\Gamma$ [cf.~Fig.~\ref{fig1}(b)]. Due to the existence of the drive field on the qubit, the component of $|g0\rangle$ will be excited to $|e0\rangle$ again. Then, the process (i.e., $|g0\rangle \rightarrow \{|\Psi_-\rangle,\, |\Psi_+\rangle\}\rightarrow (|\Psi_-\rangle,\, |g0\rangle)$) is repeated. For each cycle, the bright-state component is discarded, whereas the dark-state component is preserved and its population grows. In the long-time limit, the system will evolve from the ground state $|g0\rangle$ to the Bell state $|\Psi_-\rangle$, which is time-independent and thus serves as a stable final state. For $\Theta=\pi$, $|\Psi_{+}\rangle$ ($=|D\rangle$) is the dark state but $|\Psi_{-}\rangle$ ($=|B\rangle$) is the bright state. In this case, the dark state $|\Psi_{+}\rangle$ can be prepared by driving the qubit [cf.~Fig.~\ref{fig1}(c)].

\begin{figure}
\includegraphics[width=0.48\textwidth]{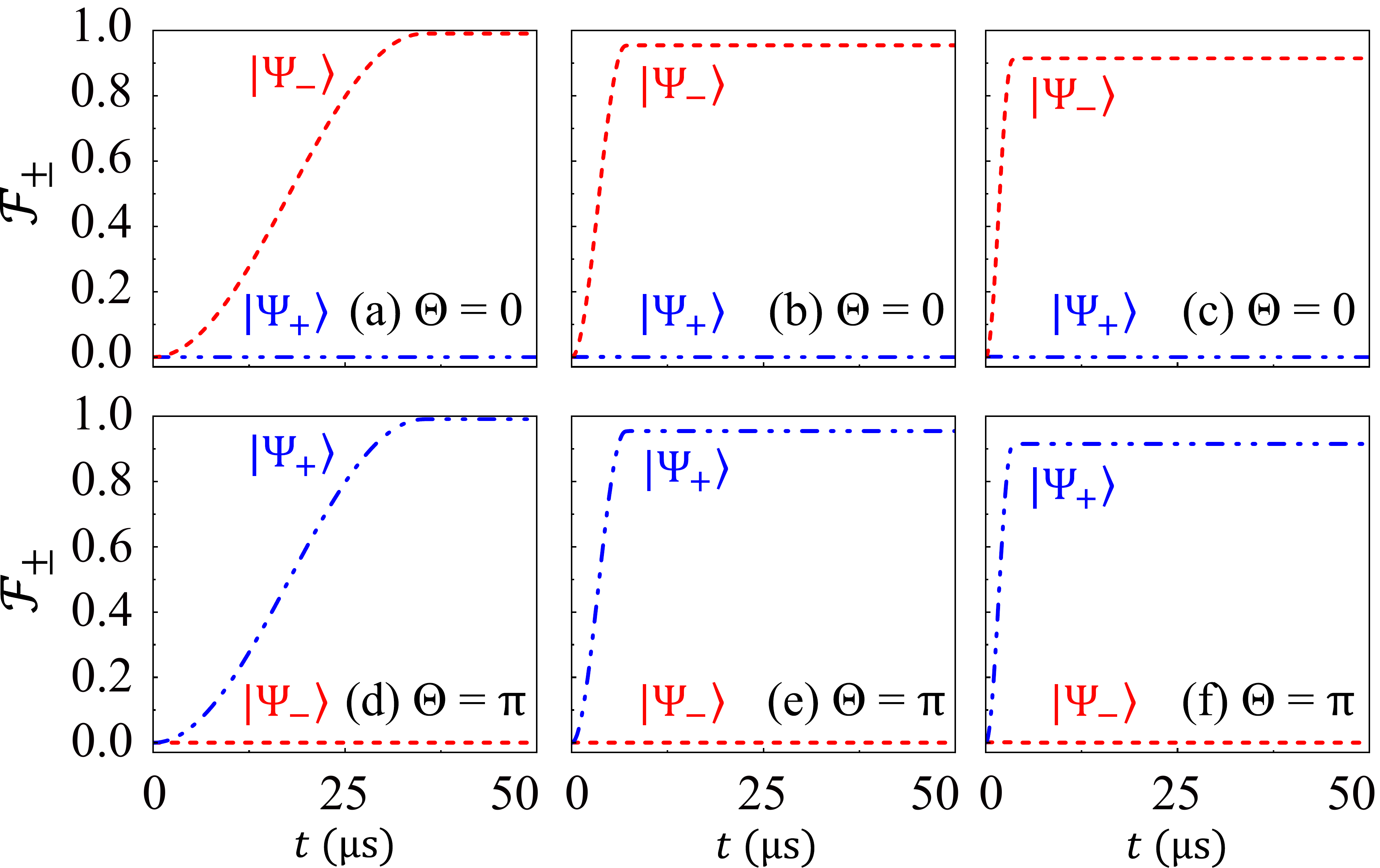}
\caption{Time evolution of the fidelity $\mathcal{F}_{\pm}$ of Bell states $|\Psi_{\pm}\rangle$, with (a)--(c) $\Theta=0$ and (d)--(f) $\Theta=\pi$, respectively, for different pulse durations and drive strengths: (a),(d) $\Omega_d/2\pi=0.01$~MHz, $t_{0}=35.335$~$\mu$s; (b),(e) $\Omega_d/2\pi=0.05$~MHz, $t_{0}=7.007$~$\mu$s; and (c),(f) $\Omega_d/2\pi=0.1$~MHz, $t_{0}=3.504$~$\mu$s. Red dashed and blue dash-dot-dot curves correspond to $|\Psi_{-}\rangle$ and $|\Psi_{+}\rangle$. Other parameters are the same as in Fig.~\ref{fig3}.}
\label{fig4}
\end{figure}

To verify the above analysis, we numerically solve the master equation~(\ref{master1}) with the Hamiltonian in Eq.~(\ref{Hamiltonian}) and plot the time evolution of the fidelity $\mathcal{F}_{\pm}={\rm Tr}(\rho |\Psi_{\pm}\rangle\langle \Psi_{\pm}|)$ of Bell states $|\Psi_{\pm}\rangle$ for $\Theta=0$ in Fig.~\ref{fig3}(a) and $\Theta=\pi$ in Fig.~\ref{fig3}(b), assuming an infinite pulse duration $t_{0}=+\infty$. When $\Theta=0$, the state $|\Psi_{-}\rangle$ ($|\Psi_{+}\rangle$) is a dark (bright) state. As shown in Fig.~\ref{fig3}(a), $\mathcal{F}_-$ increases monotonically from zero at $t=0$ to its maximum $\mathcal{F}_-^{\mathrm{(max)}}=0.990$ at $t_{\rm max}=35.34~\mu$s for $t<t_{\rm max}$ (red dashed curve), consistent with the theoretical analysis. However, the maximum fidelity remains below unity (i.e., $\mathcal{F}_-^{\mathrm{(max)}}=0.990<1$), and $\mathcal{F}_-$ exhibits oscillations in time, indicating that the prepared state $|\Psi_{-}\rangle$ is not strictly stable.
This is because $|\Psi_-\rangle$ was identified as a dark state based on the analysis without the drive field. In the presence of the drive pulse, it is no longer an ideal dark state of the full system. In contrast, the fidelity $\mathcal{F}_{+}$ of the bright state $|\Psi_{+}\rangle$ is kept nearly zero at all times (blue dash-dot-dot curve). When $\Theta=\pi$, the same drive pulse prepares the dark state $|\Psi_+\rangle$, with the maximum fidelity $\mathcal{F}_+^{\mathrm{(max)}}=0.990$ also occurring at $t_{\rm max}=35.34~\mu$s [see the blue dash-dot-dot curve in Fig.~\ref{fig3}(b)], while the bright state $|\Psi_-\rangle$ stays close to zero fidelity throughout (red dashed curve). To improve the maximum fidelity $\mathcal{F}_{-}^{\mathrm{(max)}}$ for $\Theta=0$ ($\mathcal{F}_{+}^{\mathrm{(max)}}$ for $\Theta=\pi$), one may use a weaker drive field, at the cost of a longer time $t_{\rm max}$ required to reach it [cf.~Figs.~\ref{fig3}(c) and \ref{fig3}(d)]. In the weak-drive limit ($\Omega_d \rightarrow 0$), the maximum fidelity tends to unity, but the required time $t_{\rm max}$ diverges. For comparison, Fig.~\ref{fig3}(c) [Fig.~\ref{fig3}(d)] also shows the corresponding fidelity $\mathcal{F}_+^{\mathrm{(max)}}$ ($\mathcal{F}_-^{\mathrm{(max)}}$) of bright state $|\Psi_{+}\rangle$ ($|\Psi_{-}\rangle$) at $t=t_{\rm max}$ [cf. Figs.~\ref{fig3}(a) and \ref{fig3}(b)].
For generating the stable Bell state $|\Psi_{-}\rangle$ ($|\Psi_{+}\rangle$) with fidelity $\mathcal{F}_-^{\mathrm{(max)}}$ ($\mathcal{F}_+^{\mathrm{(max)}}$) for $\Theta=0$ ($\Theta=\pi$), one can turn off the drive field upon $\mathcal{F}_-$ ($\mathcal{F}_+$) reaching its maximum. In Figs.~\ref{fig4}(a)--\ref{fig4}(c) [Figs.~\ref{fig4}(d)--\ref{fig4}(f)], we present the time evolution of the fidelity $\mathcal{F}_{\pm}$ of Bell states $|\Psi_{\pm}\rangle$ for $\Theta=0$ ($\Theta=\pi$) with several values of the Rabi frequency $\Omega_d$, where we set the pulse duration $t_0 = t_{\rm max}$. We find that in each case, $\mathcal{F}_-$ for $\Theta=0$ ($\mathcal{F}_+$ for $\Theta=\pi$) reaches its maximum $\mathcal{F}_-^{\mathrm{(max)}}$ ($\mathcal{F}_+^{\mathrm{(max)}}$) at $t=t_{\rm max}$ and remains constant thereafter, confirming the nonreciprocal generation of stable Bell states $|\Psi_-\rangle$ and $|\Psi_+\rangle$ [see Figs.~\ref{fig4}(a)--\ref{fig4}(f)]. As expected, a smaller $\Omega_d$ yields a higher $\mathcal{F}_-^{\mathrm{(max)}}$ for $\Theta=0$ ($\mathcal{F}_+^{\mathrm{(max)}}$ for $\Theta=\pi$) but requires a longer $t_{\rm max}$, cf.~Figs.~\ref{fig4}(a)--\ref{fig4}(c) [Figs.~\ref{fig4}(d)--\ref{fig4}(f)].

\begin{figure}
\includegraphics[width=0.48\textwidth]{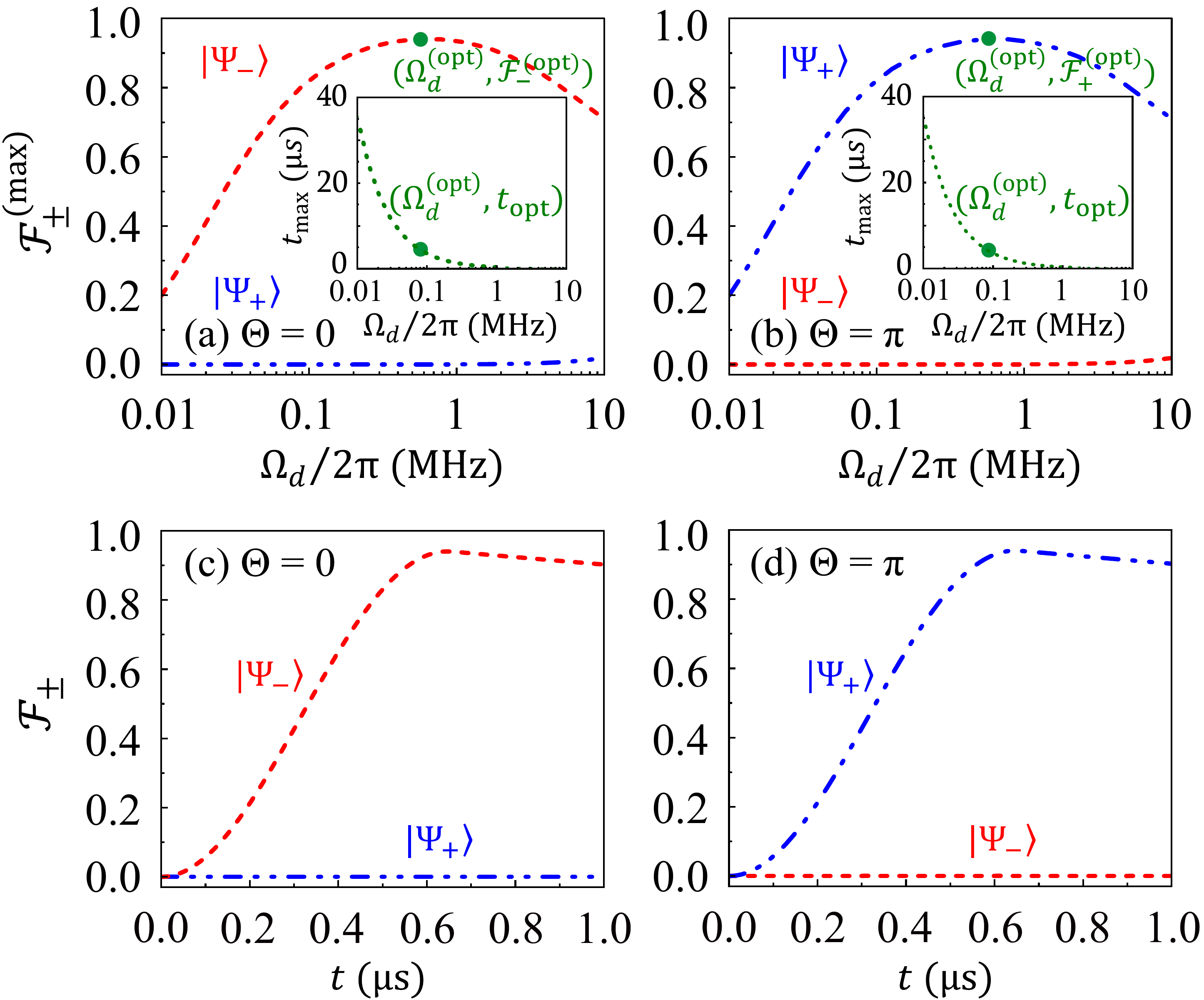}
\caption{(a),(b) Maximum fidelity $\mathcal{F}_{\pm}^{\mathrm{(max)}}$ of Bell states $|\Psi_{\pm}\rangle$ versus the Rabi frequency $\Omega_d/2\pi$ for (a) $\Theta=0$ and (b) $\Theta=\pi$, respectively, with $\gamma_m^{-1}=5~\mu$s. The corresponding time $t_{\rm max}$ required to reach $\mathcal{F}_{\pm}^{\mathrm{(max)}}$ is shown in the insets. The green dots in (a) and (b) mark the optimal fidelity $\mathcal{F}_{\pm}^{\mathrm{(opt)}}=0.940$ at $\Omega_d^{\rm (opt)}/2\pi=0.546$~MHz, and the green dots in the insets indicate the corresponding time $t_{\rm opt}=0.648~\mu$s. (c),(d) Time evolution of the fidelity $\mathcal{F}_{\pm}$ for Bell states $|\Psi_\pm\rangle$ for (c) $\Theta=0$ and (d) $\Theta=\pi$, respectively, with $\Omega_d/2\pi = 0.546$ MHz and $t_0 = 0.648~\mu$s. In all panels, red dashed curves represent $|\Psi_{-}\rangle$, and blue dash-dot-dot curves represent $|\Psi_{+}\rangle$. These results are obtained by solving the master equation (\ref{master3}) with the Hamiltonian (\ref{Hamiltonian}), the initial state $|g0\rangle$, and parameters $\gamma_q^{-1}=60~\mu$s, $\gamma_\phi^{-1}=25~\mu$s, and $\Gamma/2\pi=20$ MHz. Other parameters are the same as in Fig.~\ref{fig3}.}
\label{fig5}
\end{figure}

In any realistic experiment, the system is inevitably affected by its environment. Taking into account the dissipations of the magnon mode and the SC qubit, we modify the master equation in Eq.~(\ref{master1}) as
\begin{equation}\label{master3}
	\begin{split}
	\dot{\rho}&=-i[H,\rho]
	+\Gamma\mathcal{L}[O]\rho
	+\frac{\gamma_m}{2}\mathcal{L}[m]\rho
	+\frac{\gamma_q}{2}\mathcal{L}[\sigma^-]\rho\\
    &\quad
	+\frac{\gamma_\varphi}{2}\big(\sigma_{z}\rho\sigma_{z}-\rho\big),
	\end{split}
\end{equation}
with $\sigma_z=|e\rangle\langle e|-|g\rangle\langle g|$, where $\gamma_m$ is the relaxation rate of the magnon mode, and $\gamma_q$ ($\gamma_\varphi$) is the relaxation rate (pure dephasing rate) of the qubit. Solving the above master equation together with the Hamiltonian in Eq.~(\ref{Hamiltonian}), we plot in Figs.~\ref{fig5}(a) and \ref{fig5}(b) the maximum fidelity $\mathcal{F}_{\pm}^{\mathrm{(max)}}$ of Bell states $|\Psi_{\pm}\rangle$ versus the Rabi frequency $\Omega_d/2\pi$ for $\Theta=0$ and $\Theta=\pi$, respectively. In contrast to the ideal case without dissipation (i.e., $\gamma_m=\gamma_q=\gamma_\varphi=0$), there exists an optimal Rabi frequency $\Omega_d=\Omega_d^{\rm (opt)}$ (with $\Omega_d^{\rm (opt)}/2\pi=0.546$~MHz) at which the system reaches its optimal fidelity $\mathcal{F}_{-}=\mathcal{F}_{-}^{\mathrm{(opt)}}$ for $\Theta=0$ and $\mathcal{F}_{+}=\mathcal{F}_{+}^{\mathrm{(opt)}}$ for $\Theta=\pi$ (with $\mathcal{F}_{-}^{\mathrm{(opt)}}=\mathcal{F}_{+}^{\mathrm{(opt)}}=0.940$) at $t=t_{\rm opt}$ (with $t_{\rm opt}=0.648~\mu$s), as indicated by the green dots. This is because, when the drive is too weak, dissipation dominates the system dynamics and causes the system to relax to the ground state $|g0\rangle$. In the weak-drive limit $\Omega_d\rightarrow 0$, the system fully relaxes to the ground state, yielding $\mathcal{F}_{-}^{\mathrm{(max)}}=\mathcal{F}_{+}^{\mathrm{(max)}}=0$.
To generate the Bell state $|\Psi_-\rangle$ with $\mathcal{F}_-=\mathcal{F}_{-}^{\mathrm{(opt)}}$ for $\Theta=0$ and $|\Psi_+\rangle$ with $\mathcal{F}_+=\mathcal{F}_{+}^{\mathrm{(opt)}}$ for $\Theta=\pi$, we choose a microwave pulse with $\Omega_d=\Omega_d^{\mathrm{(opt)}}$ and $t_0=t_{\rm opt}$. As shown in Figs.~\ref{fig5}(c) and \ref{fig5}(d), the optimal fidelity $\mathcal{F}_{-}=\mathcal{F}_{-}^{\mathrm{(opt)}}$ for $\Theta=0$ and $\mathcal{F}_{+}=\mathcal{F}_{+}^{\mathrm{(opt)}}$ for $\Theta=\pi$ can be achieved. However, the resulting states $|\Psi_-\rangle$ and $|\Psi_+\rangle$ are not stable, and their lifetimes are limited by the system dissipations.

\begin{figure}
\includegraphics[width=0.48\textwidth]{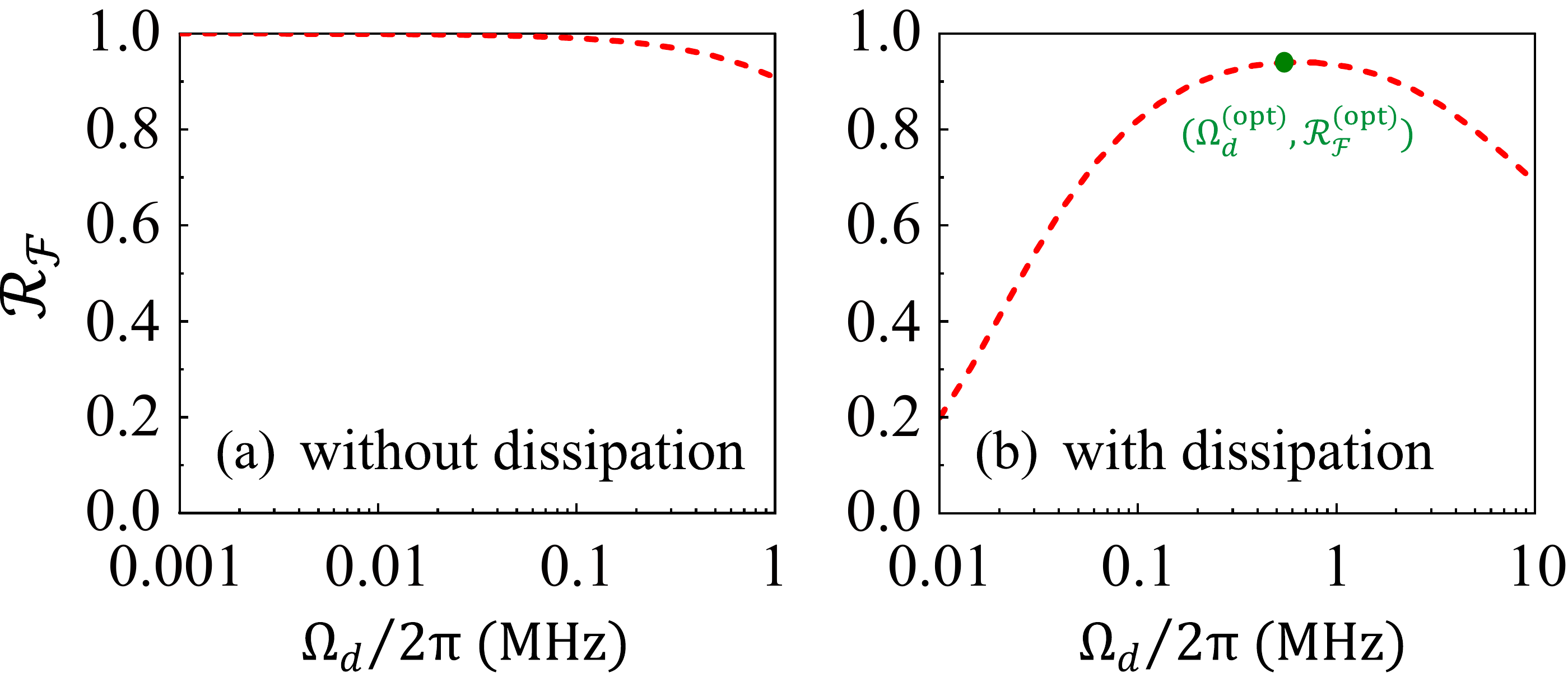}
\caption{(a) Dissipationless and (b) dissipative cases: nonreciprocal rate $\mathcal{R}_{\mathcal{F}}$ versus the Rabi frequency $\Omega_d$. The parameters used in (a) [(b)] are the same as in Fig.~\ref{fig3}(c) [Fig.~\ref{fig5}(a)].}
\label{fig6}
\end{figure}

\begin{figure}
\includegraphics[width=0.48\textwidth]{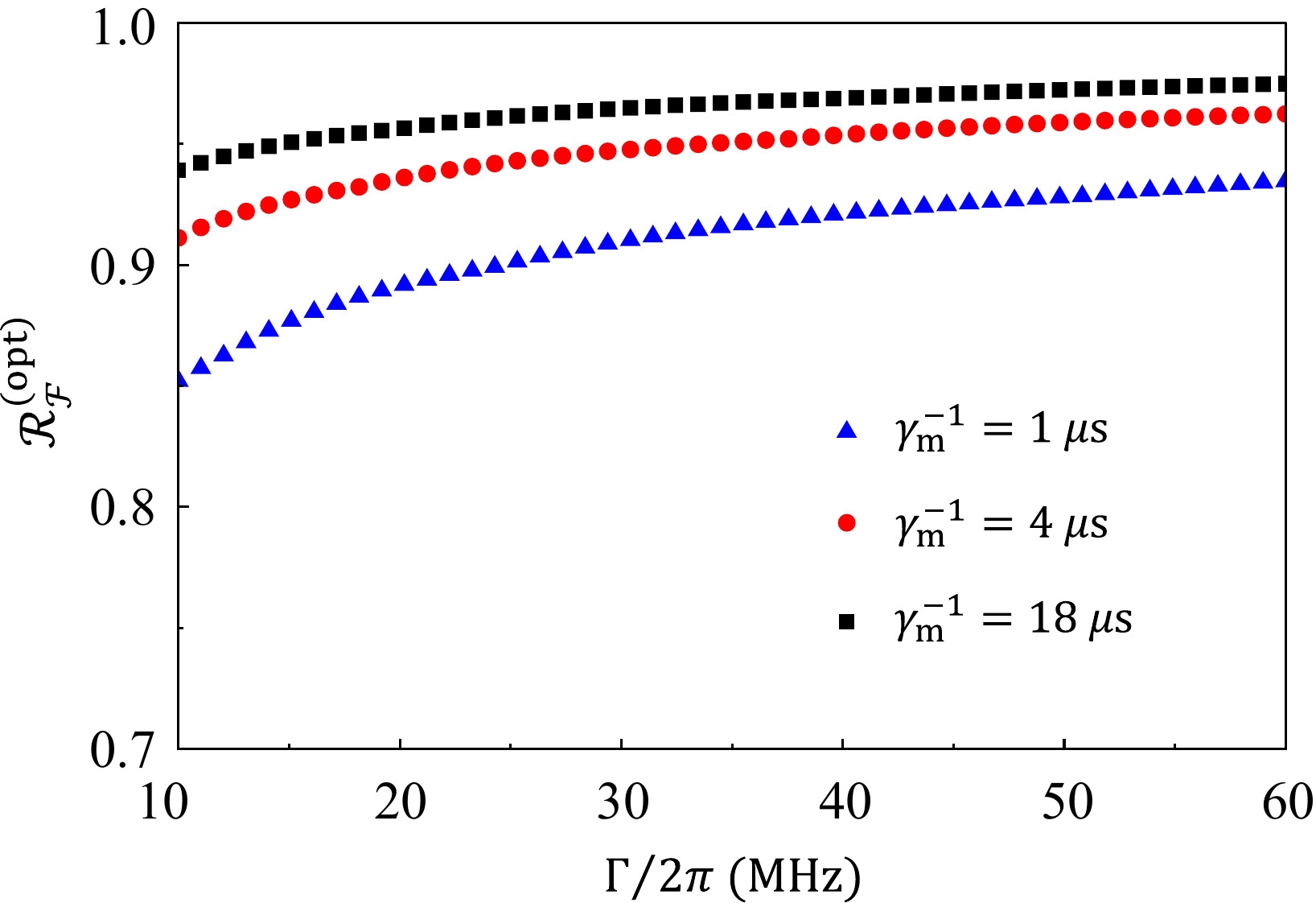}
\caption{Optimum nonreciprocal rate $\mathcal{R}_{\mathcal{F}}^{\mathrm{(opt)}}$ versus the cooperative decay rate $\Gamma/2\pi$ for different values of the magnon lifetime $\gamma_m^{-1}$. Here the blue triangles, red dots, and  black squares correspond to $\gamma_m^{-1}=1$, $4$, and $18~\mu$s, respectively. Other parameters are the same as in Fig.~\ref{fig5}(a).}
\label{fig7}
\end{figure}

To quantify the nonreciprocity of Bell-state generation, we introduce a fidelity-based nonreciprocal rate, defined as~\cite{Li24}
\begin{equation}\label{}
\mathcal{R}_{\mathcal{F}}=\left|\mathcal{F}_{-}^{\rm (max)}(\Theta=0)-\mathcal{F}_{-}^{\rm (max)}(\Theta=\pi)\right|,
\end{equation}
which satisfies $0 \leq \mathcal{R}_{\mathcal{F}} \leq 1$. Equivalently, $\mathcal{R}_{\mathcal{F}}$ can also be written as $\mathcal{R}_{\mathcal{F}}=\left|\mathcal{F}_{+}^{\rm (max)}(\Theta=0)-\mathcal{F}_{+}^{\rm (max)}(\Theta=\pi)\right|$, since the two expressions are numerically identical: $|\mathcal{F}_{-}^{\rm (max)}(\Theta=0)-\mathcal{F}_{-}^{\rm (max)}(\Theta=\pi)|
=|\mathcal{F}_{+}^{\rm (max)}(\Theta=0)-\mathcal{F}_{+}^{\rm (max)}(\Theta=\pi)|$. In general, a larger $\mathcal{R}_{\mathcal{F}}$ indicates stronger nonreciprocity. The ideal case of nonreciprocal Bell-state generation corresponds to $\mathcal{R}_{\mathcal{F}}=1$, while $\mathcal{R}_{\mathcal{F}}=0$ corresponds to the symmetric limit [i.e., $\mathcal{F}_{\pm}(\Theta=0)=\mathcal{F}_{\pm}(\Theta=\pi)$], indicating the vanishing of nonreciprocity. In the absence of system dissipation, the nonreciprocal rate $\mathcal{R}_{\mathcal{F}}$ increases monotonically as the drive strength $\Omega_d$ decreases and reaches the ideal value $\mathcal{R}_{\mathcal{F}}=1$ in the weak-drive limit $\Omega_d\rightarrow 0$ [
see Fig.~\ref{fig6}(a)]. When dissipation is included, however, $\mathcal{R}_{\mathcal{F}}$ exhibits qualitatively different behavior [cf. Figs.~\ref{fig6}(b) and \ref{fig6}(a)]. It first increases with the drive strength, reaches its optimum $\mathcal{R}_{\mathcal{F}}^{\rm (opt)}=0.940$ at $\Omega_d^{\mathrm{(opt)}}/2\pi=0.546$~MHz, and then decreases as the drive strength increases. To further improve the optimal nonreciprocal rate $\mathcal{R}_{\mathcal{F}}^{\mathrm{(opt)}}$, one can increase the cooperative decay rate $\Gamma$. As shown in Fig.~\ref{fig7}, $\mathcal{R}_{\mathcal{F}}^{\mathrm{(opt)}}$ increases monotonically with $\Gamma$ for a fixed $\gamma_m^{-1}$. For example, for $\gamma_m^{-1}=1~\mu$s ($4~\mu$s, $18~\mu$s), $\mathcal{R}_{\mathcal{F}}^{\mathrm{(opt)}}$ increases from 0.852 to 0.935 (0.911 to 0.962, 0.939 to 0.975) as $\Gamma/2\pi$ varies from 10 to 60 MHz.

\section{Discussions and conclusions}\label{Discussion}

The required parameters in the present scheme are all experimentally accessible. In typical experiments, the magnon frequency lies in the 1--10 GHz range with a lifetime $\gamma_m^{-1}$ of 1--18~$\mu$s~\cite{Tabuchi15,Lachance-Quirion17,Xu23,Weng26,Serha26}, and the SC transmon qubit operates at 1--10 GHz with decoherence times $\gamma_q^{-1}$ and $\gamma_\varphi^{-1}$ of 10--100 $\mu$s~\cite{Gu17,Rigetti12,Wei20}. Drive amplitudes $\Omega_d/2\pi \lesssim 10$ MHz are also standard for SC qubits. Since the magnon, qubit, and drive frequencies are all tunable, the resonance condition $\omega_m/2\pi=\omega_q/2\pi=\omega_d/2\pi=5$ GHz can be readily satisfied. For the SC qubit, the waveguide-induced dissipation rate has been measured to be $\mu^2\Gamma/2\pi \simeq 99.5$ MHz~\cite{Mirhosseini19}. As for the magnon mode, a dissipation rate $\nu^2\Gamma/2\pi \simeq 8.5$ MHz has been achieved by coupling a 0.25-mm-diameter YIG sphere to a waveguide~\cite{Li22}; this rate can be enhanced to $68$ MHz by increasing the sphere diameter to 0.5 mm. In particular, recent experiments have demonstrated nonreciprocal microwave transmission by reversing the bias magnetic field of a YIG sphere~\cite{Qian20,Yao25}, corresponding to the $+\mathbf{B}_0$ and $-\mathbf{B}_0$ configurations in our scheme. With these state-of-the-art experimental techniques, our scheme can be implemented in practice.

In summary, we have demonstrated the nonreciprocal generation of Bell states in a hybrid qubit-magnon system. Due to the cooperative dissipation induced by the common waveguide, a dark eigenstate emerges within the single-excitation subspace. By mapping this dark eigenstate to a target Bell state and applying a tailored drive pulse, the system can be deterministically prepared in the desired state. Reversing the bias magnetic field modifies the cooperative jump operators, directing the system into different Bell states and thereby realizing nonreciprocal state preparation. We have also defined a fidelity-based nonreciprocal rate to quantify this directional control. Our results establish cooperative dissipation as a powerful resource for manipulating quantum entangled states and offer a promising route toward chiral quantum technologies~\cite{Liu25,Li24,Khandelwal24,Wu26,Xiao26}.

\section*{Acknowledgments}

This work is supported by the National Natural Science Foundation of China (Grants No. 12674436, No. 12205069, No. 12504343, and No. 12204139), the Hangzhou Normal University scientific research and innovation team project (Grant No. TD2025003), the Hangzhou Leading Youth Innovation and Entrepreneurship Team project (Grant No. TD2024005), and the Zhejiang Provincial Natural Science Foundation of China (Grant No. LQN25A040019).

\appendix

\section{Validity of the one-excitation subspace approximation in the analytical treatment}\label{Appendix-A}

\begin{figure}
\includegraphics[width=0.48\textwidth]{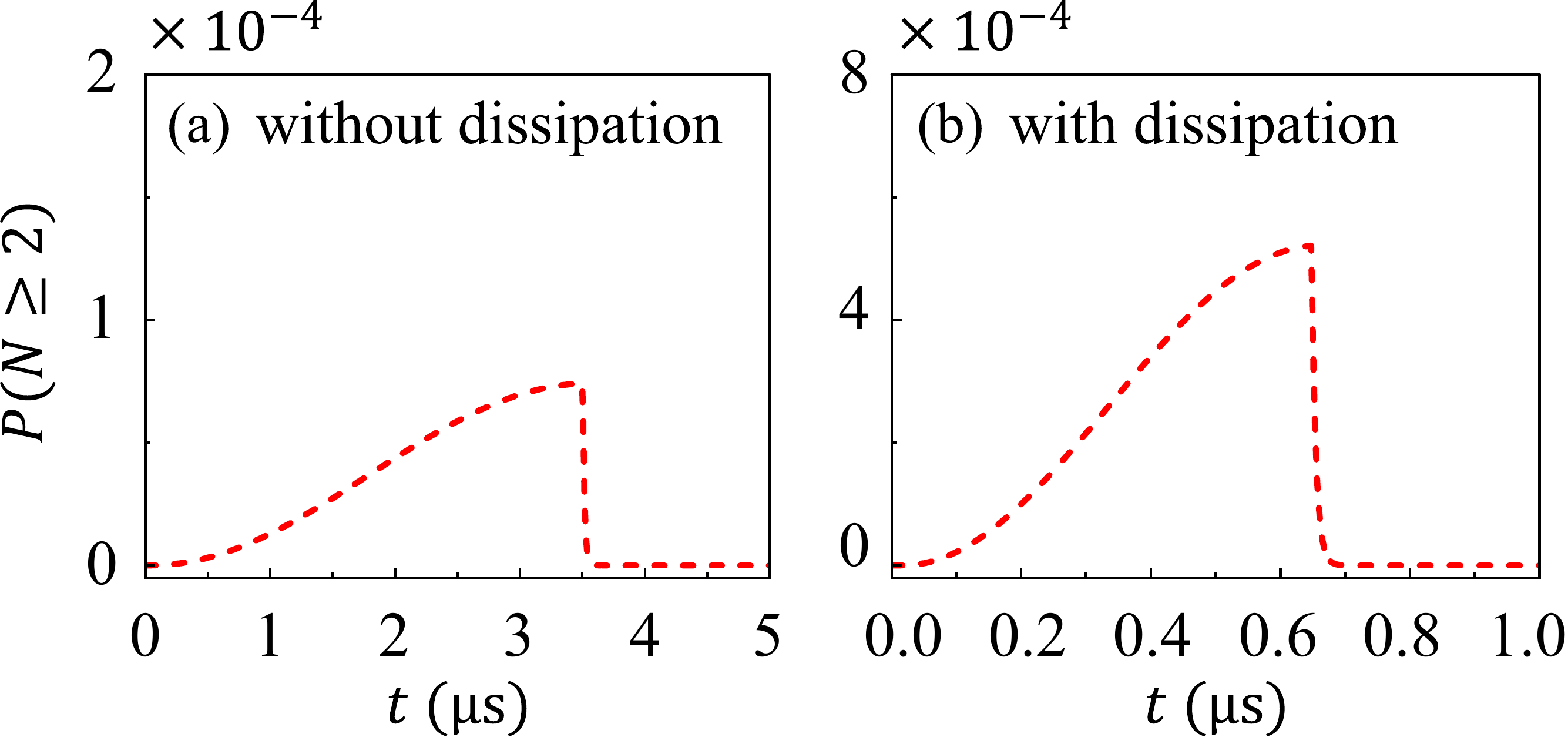}
\caption{(a),(b) Time evolution of the total occupation probability $P(N \geq 2)$ of the higher-excitation subspaces (two- and multi-excitations) for the (a) dissipationless and (b) dissipative cases. The parameters in (a) and (b) correspond to those in Figs.~\ref{fig4}(c) and \ref{fig5}(c), respectively.}
\label{fig8}
\end{figure}

In the analytical discussion of the main text, we restrict ourselves to the single-excitation subspace to illustrate the essential physical mechanism [see Eqs.~(\ref{dark-state})-(\ref{bell-state})]. In the numerical simulations, by contrast, we solve the full master equation without this restriction (cf.~Figs.~\ref{fig2}--\ref{fig7}). These numerical results can be well interpreted by this simplified model, which justifies the one-excitation approximation for the parameters considered here. The underlying physical reason is that, throughout the Bell-state generation protocol, the drive strength applied to the qubit remains much smaller than the dissipation rates of both the qubit and the magnon mode into the waveguide. As a result, the higher-excitation subspaces (two- and multi-excitations) are populated at a rate much slower than their decay into the waveguide, so that only the one-excitation subspace can retain a significant population over time due to the existence of the dark state. Furthermore, we provide numerical evidence to support the validity of the single-excitation approximation. Figures \ref{fig8}(a) and \ref{fig8}(b) show the time evolution of the total occupation probability $P(N \geq 2)$ of the higher-excitation subspaces for the dissipationless and dissipative cases, respectively. The corresponding parameters used in Fig.~\ref{fig8}(a) [Fig.~\ref{fig8}(b)] are the same as those in the Bell-state generation protocols in Fig.~\ref{fig4}(c) [Fig.~\ref{fig5}(c)]. Here $P(N \geq 2)$ is defined as $P(N \geq 2)=1-P_{\rm g0}-P_{\rm g1}-P_{\rm e0}$, with $P_j=\mathrm{Tr}(\rho |j\rangle\langle j|)$ for $j=g0, g1, e0$. The numerical results indicate that, in both cases, the population in the higher-excitation subspaces remains below $5.2 \times 10^{-4}$ throughout the entire time evolution. This confirms that the one-excitation approximation is well justified for the parameters considered in this work.

\end{document}